\documentclass[11pt]{article}
\usepackage{amsmath}

\usepackage{times}

\usepackage{moreverb}  %

\begin{document} \sloppy
\pagenumbering{roman}
\setcounter{page}{1}
\thispagestyle{empty}
\begin{center}
Argonne National Laboratory \\
9700 South Cass Avenue\\
Argonne, IL 60439

\vspace{.2in}
\rule{1.5in}{.01in}\\ [1ex]
ANL/MCS-TM-264 \\
\rule{1.5in}{.01in}

\vspace{1in}
{\Large\bf Mace4 Reference Manual and Guide}

\vspace{.2in}
by \\ [3ex]

{\Large\it William McCune}\\
\thispagestyle{empty}

\vspace{1.5in}
Mathematics and Computer Science Division

\bigskip

Technical Memorandum No. 264

\vspace{1in}
August 2003
\end{center}

\vfill
\noindent
\emph{
This work was supported by the Mathematical, Information, and Computational 
Sciences Division subprogram of the Office of Advanced Scientific Computing 
Research, Office of Science, U.S. Department of Energy, under Contract 
W-31-109-ENG-38.}
\newpage
\noindent
Argonne National Laboratory, with facilities in the states of Illinois
and Idaho, is owned by the United States Government and operated by The
University of Chicago under the provisions of a contract with the
Department of Energy.

\vspace{2in}

\begin{center}
{\bf DISCLAIMER}
\end{center}

\noindent
This report was prepared as an account of work sponsored by an agency
of the United States Government.  Neither the United States Government
nor any agency thereof, nor The University of Chicago, nor any of
their employees or officers, makes any warranty, express or implied,
or assumes any legal liability or responsibility for the accuracy,
completeness, or usefulness of any information, apparatus, product, or
process disclosed, or represents that its use would not infringe
privately-owned rights.  Reference herein to any specific commercial
product, process, or service by trade name, trademark, manufacturer,
or otherwise, does not necessarily constitute or imply its
endorsement, recommendation, or favoring by the United States
Government or any agency thereof.  The views and opinions of document
authors expressed herein do not necessarily state or reflect those of
the United States Government or any agency thereof.
\newpage
  \pagestyle{plain}
  \tableofcontents
\newpage

\pagenumbering{arabic}
\setcounter{page}{1}
\title{Mace4 Reference Manual and Guide}

\author{\emph{William McCune}}

\date{}

\maketitle

\addcontentsline{toc}{section}{Abstract}
\begin{abstract}
Mace4 is a program that searches for finite models of first-order
formulas.  For a given domain size, all instances of the formulas over
the domain are constructed.  The result is a set of ground clauses
with equality.  Then, a decision procedure based on ground equational
rewriting is applied.  If satisfiability is detected, one or more models
are printed.  Mace4 is a useful complement to first-order theorem provers,
with the prover searching for proofs and Mace4 looking for countermodels,
and it is useful for work on finite algebras.
Mace4 performs better on equational problems than did our previous
model-searching program Mace2.
\end{abstract}

\section{Introduction}

First-order satisfiability is a difficult problem---it's not even
semidecidable.  Fortunately, many of the satisfiable statements
that arise in practice have finite models, and those are what Mace4
looks for.  Most of the applications that drove the development of
Mace4 are equational, and that is the type of problem on which
Mace4 performs well.  So if you need help finding small finite
algebras, Mace4 is a good candidate.

Most of our work over the past few decades has been in automated
theorem proving, that is, searching for proofs.  In 1994 I attended
a workshop on spectra of finite quasigroups,
and that got me interested in searching for finite models of first-order
statements.  Using ideas from Mark Stickel, Hantao Zhang, John Slaney,
and others, I wrote the first version of
Mace and used it to work on several quasigroup problems \cite{dp-quasi}.
The early versions of Mace (up through version 2.*) \cite{mace2}, work 
quite differently from the one described in this report.
The early versions transform the problem into a purely
propositional (SAT) problem
and then apply a Davis-Putnam-Loveland-Logemann (DPLL)
procedure.  That method works well in many
cases, but if the original problem has clauses with many variables
or deeply nested terms, the transformation process blows up, and
only very small models can be sought.

For a given domain size, the problematic transformation
has two stages (which can be done in either order):
(1) flattening the clauses, which, in effect, pre-processes the
equality inferences, and
(2) generating all of the ground instances over the domain.
The first stage, which allows us to use a purely propositional
decision procedure (without equality), causes the blowup.

If we skip the first stage, we get a set of ground clauses
with equality, which is clearly a decidable problem.
Jian Zhang's FALCON \cite{falcon} works
in this way, with a special-purpose decision procedure.
In particular, it has an optimization, the \emph{least-number
heuristic} (LNH), to take advantage of symmetries in the problem.
Shortly after FALCON was developed, Jian Zhang and Hantao Zhang
produced SEM \cite{sem}, which inherited those key methods of
FALCON, and which has been distributed and used in several
projects.

Mace4 owes much to FALCON and SEM.  It transforms the original
problem to a ground clauses with equality and uses a special-purpose
decision procedure with the LNH.  It also uses negative inference
rules as in FALCON and SEM.

\section{A Little Motivation} \label{sec-motivation}

Suppose you wish to see a small noncommutative group.
You prepare the following input file, named \texttt{group.in}.
\begin{center}
\begin{small}
\begin{boxedverbatim}
assign(iterate_up_to, 10).  %

clauses(theory).
E * x = x.                  %
x' * x = E.                 %
(x * y) * z = x * (y * z).  %
A * B != B * A.             %
end_of_list.
\end{boxedverbatim}
\end{small}
\end{center}
\noindent
Then, you give the input to Mace4 to look for a model as follows.

\begin{small}
\begin{verbatim}
    %
\end{verbatim}
\end{small}

\noindent Mace4 succeeds, and the output file contains the following
noncommutative group of order 6.

\begin{center}
\begin{small}
\begin{boxedverbatim}
- Model 1 at 0.01 seconds -
  
  E : 0    A : 1    B : 2
  
  ' :    0 1 2 3 4 5
     ---------------
         0 1 2 4 3 5
  
  * :  | 0 1 2 3 4 5
     --+------------
     0 | 0 1 2 3 4 5
     1 | 1 0 3 2 5 4
     2 | 2 4 0 5 1 3
     3 | 3 5 1 4 0 2
     4 | 4 2 5 0 3 1
     5 | 5 3 4 1 2 0
\end{boxedverbatim}
\end{small}
\end{center}

\section{The Input File}

The input file starts with an optional set of commands
that tell Mace4 how to search.
The rest of the file contains lists of clauses or formulas
representing the theory,
that is, the clauses or formulas you wish to model.
Comments can appear on any line in the file;
they start with the character ``\texttt{\%}'' and end
at the end of the line.

\subsection{The Commands}  \label{sec-commands}

\newcommand{\spacer}{000000000000000000000000000000000000}

\newcommand{\assigner}[2]{\begin{tabbing}\spacer\=\kill\texttt{assign(#1, $n$).} \> \texttt{\% default $n=#2$}\end{tabbing}}

\newcommand{\setter}[1]{\begin{tabbing}\spacer\=\kill\texttt{set(#1).} \> \texttt{\% default set}\\ \texttt{clear(#1).}\end{tabbing}}

\newcommand{\clearer}[1]{\begin{tabbing}\spacer\=\kill\texttt{set(#1).}\\ \texttt{clear(#1).} \> \texttt{\% default clear}\end{tabbing}}

\newcommand{\other}[1]{\begin{tabbing}\texttt{#1}\end{tabbing}}

Mace4 has only a few types of command.  The \texttt{set} and
\texttt{clear} commands apply to flags, the \texttt{assign}
commands apply to integer-valued parameters, and the \texttt{op}
commands are syntactic sugar, allowing expressions to be written
in more human-friendly ways.

\subsubsection{Basic Commands}

\assigner{domain\_size}{2}
\begin{quote}
Search for models of size $n$.
See the following command if you wish for Mace4 to iterate
through larger domain sizes.
The default value of \texttt{domain\_size} is $n=2$.
\end{quote}

\assigner{iterate\_up\_to}{0}
\begin{quote}
If $n >$ \texttt{domain\_size}, Mace4 will iterate through domain
sizes, restarting the search each time.  For example if
\texttt{domain\_size}=4 and \texttt{iterate\_up\_to}=6,
Mace4 will try 4, then 5, then 6, until those sizes are
exhausted or until some other limit such as \texttt{max\_models}
applies.
\end{quote}

\setter{print\_models}
\begin{quote}
If this flag is set, all found models are printed to the output file.
The format of the models depends on the state of the flag
\texttt{print\_models\_portable}.
\end{quote}

\clearer{print\_models\_portable}
\begin{quote}
If this flag is set,
models are printed to the output file in \emph{portable format},
which means that they can be read by various other programs, for example,
a program that takes a list of models and removes isomorphic ones,
or a program that uses models to filter clauses.
\end{quote}

\assigner{max\_models}{1}
\begin{quote}
The search will terminate when the $n$th model is found.
(Typically, many of the models are isomorphic; see the
program \texttt{isofilter} in Section \ref{sec-aux}.)
If domain-size iteration is occuring, the
model count carries over from each size to the next.
The value $n= -1$ means that there is no limit.
\end{quote}

\assigner{max\_seconds}{-1}
\begin{quote}
The search will terminate if it reaches $n$ seconds
(user CPU time on UNIX-like systems).
If domain-size iteration is occuring, the
timer carries over from each size to the next.
The value $n= -1$ means that there is no limit.
\end{quote}

\assigner{max\_megs}{192}
\begin{quote}
The search will terminate if it tries to dynamically
allocate (malloc) more than $n$ megabytes of memory.
(The entire Mace4 process can consume somewhat more than
\texttt{max\_megs} megabytes.)
\end{quote}

\clearer{prolog\_style\_variables}
\begin{quote}
A rule is needed to distinguish variables from constants in clauses.
Normally, variables start with (lower case) \texttt{u} -- \texttt{z},
and all other nullary symbols are constants.
If this flag is set, variables start with (upper case)
\texttt{A} -- \texttt{Z}.
\end{quote}

\clearer{verbose}
\begin{quote}
If this flag is set, the output file receives information
about the search, including the initial partial model (the part of
the model that can be determined before backtracking starts)
and timing and other statistics for each domain size.  (It does
not give a trace of the backtracking, so it does not consume a lot
of file space.)
\end{quote}

\begin{tabbing}
\texttt{op(\textrm{\emph{precedence}}, \textrm{\emph{type}}, \textrm{\emph{symbol}}).}\\
\texttt{op(\textrm{\emph{precedence}}, \textrm{\emph{type}}, \textrm{\emph{list-of-symbols}}).}
\end{tabbing}
\begin{quote}
This command allows expressions to be written in infix, postfix,
and prefix forms without so many parentheses.
For example, the command
\hspace{.25cm}\texttt{op(400,infix,*)}\hspace{.25cm}
declares ``\texttt{*}'' to be an infix (binary) operation with precedence 400.
This command affects how clauses and formulas are parsed and printed only.
The accepted types for binary operations are
\texttt{infix},
\texttt{infix\_left} (left association), and
\texttt{infix\_right} (right association).
The accepted types for unary operations are
\texttt{prefix} and
\texttt{postfix}.
The precedence should be in the range 1--998.
Symbols with lower precedence bind more tightly.
Several operations are predeclared---see the table in
Section \ref{sec-clauses}.
\end{quote}

\subsubsection{Advanced Commands}

The commands in this sections can be used to fine-tune the search.

\setter{lnh}
\begin{quote}
This flag says to use the \emph{least number heuristic} (LNH).
It is nearly always recommended.
See Section \ref{sec-lnh}.
\end{quote}

\assigner{selection\_order}{2}
\begin{quote}
This parameter determines the set of cells that are
candidates for selection.  See Section \ref{sec-cell-selection}.
If $n=0$, all open cells are considered.
If $n=1$, \emph{concentric order} is used.
If $n=2$, \emph{concentric-band order} is used.
\end{quote}

\assigner{selection\_measure}{4}
\begin{quote}
Given a set of cells, this parameter determines which of
those cells is selected for assignment.
See Section \ref{sec-cell-selection}.
If $n=0$, select the first candidate cell.
If $n=1$, select the candidate with the greatest number of occurrences
in the current set of (ground) clauses.
If $n=2$, select the candidate that would cause the greatest number
of propagations.
If $n=3$, select the candidate that would cause the greatest number
of contradictions.
If $n=4$, select the candidate with the fewest possible values.
\end{quote}

\setter{negprop}
\begin{quote}
This flag enables or disables the \emph{negative propagation}
inference rules, which derive negated equations.
These rules are not required for completeness of the search,
but they nearly always help.
Four types of negative propagation are 
enabled or disabled by the following four parameters.
For any negative propagation to occur, \texttt{negprop} must be set.
\end{quote}

\setter{neg\_assign}
\begin{quote}
If this flag is set, negative propagation is triggered by \emph{assignments}
(i.e., when a cell is given a value), for example, $f(2,3)=4$.
\end{quote}

\setter{neg\_assign\_near}
\begin{quote}
If this flag is set, negative propagation is triggered by
\emph{near assignments}, for example, $f(2,g(3))=4$.
\end{quote}

\setter{neg\_elim}
\begin{quote}
If this flag is set, negative propagation is triggered by
\emph{eliminations}, for example, $f(2,3) \neq 4$.
\end{quote}

\setter{neg\_elim\_near}
\begin{quote}
If this flag is set, negative propagation is triggered by
\emph{near eliminations}, for example, $f(2,g(3)) \neq 4$.
\end{quote}

\clearer{trace}
\begin{quote}
If this flag is set, detailed information about the search,
including a trace of all assignments and backtracking, is
printed to the standard output.  \emph{This flag causes a lot of
output, so it should be used only on small searches}.
\end{quote}

\subsection{The Clauses and Formulas}

The theory to be modeled can be specified with first-order
clauses, first-order formulas, or a combination of the two.
There can be any number of clause and formula lists, given
as in the following (silly) example.
\begin{small}
\begin{verbatim}
    clauses(theory_part_1).   %
    (x * y) * z = x * (y * z).
    end_of_list.

    formulas(theory_part_2).  %
    all x all y (x * y = y * x).
    end_of_list.
\end{verbatim}
\end{small}
The names of the lists need not be ``\texttt{theory\_part\_1}'' and
``\texttt{theory\_part\_2}''.
Any names can be used.
When formulas are given, they are immediately transformed
(including Skolemization) to clauses by a straightforward procedure.

The clause and formula languages accepted by Mace4 are determined
by a library of automated deduction routines that is still
evolving (LADR \cite{ladr-www}),
so they are likely to change in future releases of Mace4.
Therefore we do not give a full definition here.
Instead we list some key points and examples, and we refer the reader
to the sample input files that accompany the Mace4 distribution packages.
A more formal description of the languages may appear on the Mace4
Web page \cite{mace4-www}.

Parsing and printing properties of binary and unary symbols
can be declared, allowing infix, prefix, and postfix notation.
These declarations are syntactic sugar only---they have no logical
significance.  The user can declare such properties with
the \texttt{op} command, and several are predeclared as shown in
the following table.

\begin{center}
\begin{small}
\begin{boxedverbatim}
op(800, infix,       ->).  %
op(800, infix,       <-).  %
op(800, infix,      <->).  %
op(790, infix_right,  |).  %
op(780, infix_right,  &).  %
op(300, prefix,       ~).  %
op(700, infix,        =).  %
op(700, infix,       !=).  %
op(500, infix,        +).  %
op(400, infix,        *).  %
op(300, prefix,       -).  %
op(300, postfix,      ').  %
\end{boxedverbatim}
\end{small}\\[.2cm]
Predeclared Operations
\end{center}

Whitespace (spaces,newlines,returns,tabs,vertical-tabs,formfeeds)
is allowed just about anywhere.
When in doubt about how much white
space and how many parentheses to use, be conservative and
observe how Mace4 echoes your input.  Mace4 frequently prints
more whitespace than necessary.

Negated equality can be abbreviated with the symbol ``\verb:!=:''.

\subsubsection{Domain Elements} \label{sec-domain}

If the input clauses or formulas contain constants that are
natural numbers, $\{0,1,2,3,\ldots\}$, such constants are always
interpreted as members of the domain.  For example, one can use 1 and 0
as the top and bottom of a lattice, and the corresponding members
of the domain are assigned to them.

When using domain elements (i.e., natural number constants) in the input,
it is best to use the smallest ones---doing otherwise
defeats the purpose of the least-number heuristic.
\emph{If the input contains a
domain element that is greater than or equal
to the specified domain size, a fatal error occurs.}

\subsubsection{Clauses} \label{sec-clauses}

A clause is a disjunction of literals, terminated with a
period.  A literal is
either an atomic formula or the negation of an atomic formula.
The disjunction symbol is \verb:|:, and the negation symbol is \verb:~:.

All quantification in clauses is universal and implicit.
Variables are distinguished from constants by the following rule.
Ordinarily, variables start with (lower case) \texttt{u}--\texttt{z}.
If the flag \texttt{prolog\_style\_variables} is set,
variables start with (upper case) \texttt{A}--\texttt{Z}.

\begin{center}
\begin{small}
\begin{boxedverbatim}
p(x) | ~q(a) | r(x,b).        %
|(p(x),|(~(q(a)),r(x,b))).    %
x*y!=z | x*w!=z | y=w.        %
x * (y + z) = x * y + x * z.  %
x' ' = x.                     %
\end{boxedverbatim}
\end{small}\\[.2cm]
Examples of Clauses, Assuming the Default Operator Declarations
\end{center}

\subsubsection{Formulas}

All quantification in formulas is explicit, and the top level of
a formula does not have free variables.
Therefore, no rules are needed to distinguish variables from
constants---a nullary symbol is a variable if and only if it is
bound by a quantifier.  For example, the symbol \texttt{e} can be
a variable.  To prevent a common error, however, we do not allow
constants that might be misinterpreted by the user as variables.
For example, the string \verb:(all x p(x,y)): is not accepted
as a formula, because the constant \verb:y: looks like a variable.

\begin{itemize}
\item An atomic formula is a formula.
\item If \verb:F: is a formula, then \verb:(~F): is a formula.
\item If \verb:F: and \verb:G: are formulas, then
\verb:(F & G):,
\verb:(F | G):,
\verb:(F -> G):,
\verb:(F <- G):, and
\verb:(F <-> G):
are formulas.
\item If \verb:x: is a symbol and \verb:F: is a formula,
then \verb:(all x F): and \verb:(exists x F):
are formulas.
\end{itemize}

The default operator declarations allow many parentheses to be dropped.
Also, a special rule allows parentheses to be dropped for
sequences of quantifiers.  For example, the following two strings
represent the same formula.

\begin{small}
\begin{verbatim}
    all x all y exists z (p(x,y,z) <-> q(x) & ~r(y) | s(z)).
    (all x (all y (exists z ((p(x,y,z) <-> (q(x) & (~r(y))) | s(z)))))).
\end{verbatim}
\end{small}

\section{Running Mace4}

The basic way to call Mace4 is shown in the
following example.

\begin{small}
\begin{verbatim}
    %
\end{verbatim}
\end{small}
Mace4 also accepts command-line arguments corresponding
to the options described in Section \ref{sec-commands}.
Command-line options override the corresponding settings
in the input file.  To see the correspondence between the
options and the command-line arguments, run the following
command.

\begin{small}
\begin{verbatim}
    %
\end{verbatim}
\end{small}
Assume the command-line argument \texttt{-n} corresponds to
the parameter \texttt{domain\_size}, and \texttt{-m}
corresponds to \texttt{max\_models}.
If the input file contains the command
\texttt{assign(iterate\_up\_to,10)},
the command
\begin{small}
\begin{verbatim}
    %
\end{verbatim}
\end{small}
tells Mace4 to search for up to 20 models of sizes 8, 9, and 10.

\subsection{Theorem Prover Compatibility Mode}

One of the basic ways to use Mace4 is as a complement to
a theorem prover, with the prover searching for a proof
and Mace4 looking for a counterexample.  In order to be
able to use the same input file with both programs, Mace4
has a mode in which it will allow (and ignore) commands
and lists intended for other programs.
(Ordinarily an unrecognized command or list causes a fatal error.)

The command-line option \texttt{-c} tells Mace4 to allow
unrecognized \texttt{set}, \texttt{clear}, and \texttt{assign} commands
and to allow unrecognized lists of objects.

Mace4 is not compatible with Otter \cite{otter3} in this way.
However, a script can help in transforming
Otter (and Mace2) input files for use with Mace4
(see Section \ref{sec-aux}).

\subsection{Exit Codes}

When a Mace4 process terminates, it returns an exit code.
If Mace4 is called from another program,
say a shell or Perl script, that program can use Mace4's
exit code to decide what do do next.  The exit codes are as follows.
\begin{enumerate}
\item[0:]
The specified number of models (\texttt{max\_models}, default 1) was found.
\item[1:]
A fatal error occurred.  This is usually caused by
an error in the input file,
the memory limit (\texttt{max\_megs}, default 192), or
a bug in Mace4.
\item[2:]
The search completed without finding any models.  That is,
there are no models within the given constraints.
\item[3:]
Some models were found, but the search completed before
\texttt{max\_models} models were found.
\item[4:]
Some models were found, but the
time limit (\texttt{max\_seconds}, default $\infty$) terminated the search
before \texttt{max\_models} models were found.
\item[5:]
The time limit (\texttt{max\_seconds}) terminated the search
before any models were found.
\end{enumerate}

Here is an example of a Perl program that calls Mace4.
It takes a stream of clauses and for each, it calls
Mace4 to look for a noncommutative model up through size four.
If none is found, the clause is printed.

\begin{center}
\begin{small}
\begin{boxedverbatim}
#!/usr/bin/perl -w
$mace4 = "/home/mccune/bin/mace4";
$input = "/tmp/mace$$";
$unsatisfiable_exit = 2;
while ($equation = <STDIN>) {
  open(FH, ">$input") || die "Cannot open file $input";
  print FH "clauses(theory). $equation f(0,1)!=f(1,0). end_of_list.\n";
  close(FH);
  $rc = system("$mace4 -N4 < $input > /dev/null 2> /dev/null");
  $rc = $rc / 256;	# This gets the actual exit code.
  if ($rc == $unsatisfiable_exit) { print $equation; }
}
system("/bin/rm $input");
\end{boxedverbatim}
\end{small}
\end{center}

\section{How Mace4 Works}

This section describes how Mace4 works for a fixed domain size,
say $n$.  The members of the domain are always named $\{0,\ldots,n-1\}$.
If Mace4 is iterating through domain sizes, this section applies,
separately, to each domain size.

First, tables for the function and predicate symbols are set up, and
all ground instances of the input clauses (over the domain) are
generated.  Then, in a systematic way, a recursive backtracking
procedure fills in the cells of the tables and uses the ground clauses
to propagate the effects of the assignments.  When contradictions
(dead ends) are encountered, backtracking occurs, the propagations and
assignments are undone, and other assignments are attempted.  If all
the tables become full, with no contradictions, a model has been
found.  That is, we have an interpretation in which all of the input
clauses are true.

\subsection{Initialization}

Mace4 starts by allocating a table for each function and
predicate symbol.
Constants (which are function symbols of arity 0) are single cells,
function symbols of arity 2 get an $n \times n$ table, and so on.
The range of values for function symbols will be members of the domain
$[0,\ldots,n-1]$, and values for predicate symbols will be
0 and 1, representing the truth values.

Then, for each input clause, all ground instances over the domain are
generated.  If a clause has $v$ variables, it has $n^v$ instances.  A
variable \emph{max\_constrained} contains the value of the
maximum constrained value and is used for the \emph{least number
heuristic} (see Section \ref{sec-lnh}).
If any domain values occur in the input
clauses, \emph{max\_constrained} is initialized to the greatest domain
value in the input; otherwise it is initialized to -1.

In many cases, the input clauses allow us to start filling
in the tables before the search begins.  For example, if the
input says that a binary operation is idempotent, say $f(x,x)=x$,
then the $n$ cells on the diagonal of table $f$ can be filled.
This can be done by simply calling the assignment and propagation
routines described below.  All of these are fixed assignments
which never need to be undone.

At this point, the initialization is complete, and the initial partial
model (possibly empty) may be printed to the output file.
Now, we can start the search.

\subsection{Search}

Here is the recursive backtracking search procedure, roughly stated.
\begin{center}
\begin{small}
\begin{boxedverbatim}
  procedure search:
  {
      cell = select_cell();
      top = last_value_to_consider();
      foreach i (0 ... top)
      {
          ok = assign_and_propagate(cell,i); 
          if (ok)
          {
              search();
              undo_assignments();
          }
      }
  }
\end{boxedverbatim}
\end{small}
\end{center}
The important components of the search are cell selection,
determining which values need to be considered for assignment to
the selected cell, and propagating the assignments.
Note that the same pseudo-code could be
used to represent a Davis-Putnam-Loveland-Logemann (DPLL) propositional
satisfiability procedure.  In that case, \texttt{select\_cell()}
would select a propositional variable, and
\texttt{top} is always 1.

\subsection{Cell Selection} \label{sec-cell-selection}

Cell selection is divided into two stages.
The first, controlled by the parameter \texttt{selection\_order},
determines a set of candidate cells.  The second stage,
controlled by the parameter \texttt{selection\_measure}, picks
one of the candidates cells.

\paragraph{Candidates Cells.}  Three methods are available,
specified by the parameter \texttt{selection\_order}.
\begin{itemize}
\item[0.] The \emph{linear order}.  All open cells are candidates.
This order usually defeats the purpose of the least number heuristic.
\item[1.] The \emph{concentric order}.  
Let $i$ be the smallest maximum index of an open cell;
then all cells with maximum index $i$ are candidates.
This method causes the maximum constrained value to be
increased only when necessary.
\item[2.] The \emph{concentric-band order}.  
All cells with maximum index less than or equal to the
current maximum constrained value are candidates.
If there is none, we revert to concentric order.
This method usually gives more candidates than the concentric order,
but it still keeps maximum constrained value as low as possible.
\end{itemize}

\paragraph{Selecting a Candidate.}  Four methods are available,
specified by the parameter \texttt{selection\_measure}.
\begin{itemize}
\item[0.]
Select the first candidate.
\item[1.]
Select the candidate with the greatest number of occurrences
in the current set of (ground) clauses.
Recall that each cell is associated with a term, e.g., $f(2,3)$.
The motivation for this
measure is that the assignment will have
the greatest initial effect (ignoring propagation) in simplifying
the clauses.
\item[2.] 
Select the candidate that would cause the greatest number
of propagations.  This is done as follows.
For each cell being considered, all assignments
and subsequent propagations are done (and undone), and the
total number of propagations is counted; this is an expensive
lookahead operation with the motivation of filling the tables
as soon as possible.
\item[3.]
Select the candidate that would cause the greatest number
of contradictions; for each cell being considered, all assignments
and subsequent propagations are done (and undone), and the number
of assignments that lead directly to contradictions are counted;
this is an expensive lookahead operation with the
motivation of cutting off paths as soon as possible.
\item[4.]
Select the candidate with the smallest number of possible values.
Each cell has a list of possible values (see Section \ref{sec-theory}).
As the search progresses, negated equalities can be derived.
If, for example, $f(2,3)\neq 4$ is derived, the value 4 is removed
from the possible-values list for cell $f(2,3)$.  (When
all but one value is eliminated, an assignment can be made.)
The primary benefit of negative propagation (see Section \ref{sec-negprop})
is that it derives negated equalities that are used for this
purpose.

\end{itemize}

\subsection{Cell Assignment} \label{sec-lnh}

Once we select an open cell, we have to determine the set of values
to be considered for assignment to the cell.  If it is a
Boolean-valued cell, we always try both 0 and 1.  Otherwise,
if the \texttt{LNH} (least-number heuristic) flag is clear,
we try all members of the domain $[0,\ldots,n-1]$.

The \emph{least number heuristic} (LNH) was first
used in Falcon \cite{falcon} and later in SEM \cite{sem}.
It eliminates some of the isomorphism in the search.
The basic idea is that at a given point in the search,
all of the domain values, say $[0,\ldots,n-1]$,
are partitioned into the constrained values $[0,\ldots,i]$ and
the unconstrained values $[i+1,\ldots,n-1]$.  The unconstrained
values are all symmetric---if any one of them is tried, the others
need not be.  We try the least of them.  That is, when faced with
an empty cell, we try $[0,\ldots,i+1]$.  A value becomes constrained
when it is assigned to a cell or when it becomes the index of a cell
that is assigned to.  (Subsequent assignments during unit propagation
do not constrain values.  Because of this distinction, we often
think of two types of assignment: by selection and by propagation.)
The LNH eliminates only some of the isomorphism in the search.
In practice, it usually has effect only at the very top levels of
the search tree.

\subsection{Positive Propagation}

When we make an assignment, say $f(2,3)=4$, positive propagation
is always applied.  All occurrences of $f(2,3)$ in the ground
clauses are rewritten to 4, and these changes can trigger other
rewrite operations by the other current assignments.
During this process, more assignments can be derived.  For example,
with equation $g(f(2,3))=5$, we rewrite once, and then we have
assignment $g(4)=5$ to propagate.  And so on.  Positive propagation
is important to the efficiency of the method; there are typically
many assignments by propagation for each assignment by selection.
Each rewriting operation and each assignment are recorded on
a stack so that they can be undone when backtracking.  In addition,
we maintain a queue of assignments to be made and propagated.

\subsection{Negative Propagation} \label{sec-negprop}

Negative propagation derives negated equalities.
The power of this feature is in eliminating possible values.
If all but one value for a cell is eliminated, an assignment
can be made.  In addition, cell selection can be based on
the least number of possible values, under the motivation that
fewer possible values causes less branching.

Suppose we know that $f(2,3)=4$ and $f(2,g(5))\neq 4$;
then we may infer $g(5)\neq 3$.  The same conclusion can be
drawn if the signs on the premises are reversed.
Examples of the four types of clause to which negative
propagation can be applied are shown in 
the following table.
\begin{center}
\begin{tabular}{ll}
Type  & Example\\
\hline
Assignment       & $f(2,3)=4$ \\
Near assignment  & $f(2,g(5))=4$ \\
Elimination      & $f(2,3)\neq 4$ \\
Near elimination & $f(2,g(5))\neq 4$ \\
\hline
\end{tabular}
\end{center}
Assignments pair with near eliminations, and eliminations pair
with near assignments.  Any of the four types can trigger
negative inferences, and there are four corresponding flags
to enable the negative inferences.  For example, if the
\texttt{neg\_elim\_near} flag is set, and if a near elimination
clause is derived, it will be used with all applicable assignment
clauses to derive new negated equalities.  Negative inferences
can cause propagation of assignments as well as eliminations
by eliminating all but one possible value of a cell.

Negative inference and propagation apply to Boolean-valued terms as well
as to ordinary terms.  If we have $P(3,4)$ and $\neg P(3,g(5))$, we can
derive $g(5)\neq 4$.  All of the rules for Boolean terms are analogous
to the rules for ordinary terms.

Negative propagation is not required for completeness
and can be disabled with the command \texttt{clear(neg\_prop)}.

\section{Implementation}

Mace4 is coded in C, and and we have tried to make it run quickly.
As in most DPLL procedures, the vast majority of the processing time is
spent propagating assignments.  In our case, that includes
both positive and negative propagation.

The data structures were designed with the goal of minimizing
the the allocation and deallocation of memory.  After the ground
clauses are constructed, no more term allocation occurs; all
of the rewriting is done by copying pointers.  Memory is
allocated and deallocated (by our own routines, not by
operating system calls) for the stack of changes that must
be undone and for the queue of tasks to be done.

Say we have a new assignment $f(1,2)=3$.  We maintain a list
of occurrences of $f(1,2)$ (constructed from the term nodes
so that it need not be allocated or deallocated); for each
occurrence, say $g(f(1,2))$, we rewrite it to $g(3)$ and
record that change on the stack.  We then update the list
of occurrences of $g(3)$ to include this new one and record
that change in the stack.  If this rewrite creates a new
assignment, elimination, near assignment, or near elimination,
we add that to the queue of tasks.

For negative inference, we have to be able to quickly locate
near assignments and near eliminations as well as ordinary
assignments and eliminations.  We index near assignments and
near eliminations with \emph{complete discrimination trees},
that is, a discrimination tree \cite{indexing} in which all
possible branches are constructed at the start of the search,
and only the leaves are updated for insertions and deletions.
The complete discrimination trees are not very big, because
the term depth is always 2.  Insertions into the discrimination
trees are recorded on the stack so that they are undone on backtracking.

Mace4 is the first released program that has been constructed with
LADR (Library of Automated Deduction Routines) \cite{ladr-www},
a library of C routines for building automated deduction tools.
LADR has been under development at Argonne, off and on, for several
years (at times LADR was also known as OPS).  The LADR features
we used involve the term data structure, parsing and printing
terms, interpretations, auxiliary data structures such as linked
lists, flag and parameter handling, and clocks.

\section{Theory and Completeness} \label{sec-theory}

For a given (finite) domain size, the existence of a model is
obviously decidable, and Mace4 is intended to be a decision procedure.
We can view the existence problem as satisfiability of a set of ground
clauses with equality.

To take this view, we start with the set of ground clauses that
Mace4 generates during initialization.
Then we add some clauses that correspond to operations that are
built in to Mace4.
First, we add $(n^2-n)/2$ clauses stating that the $n$ domain
members are distinct.
Second, for each cell we add an $n$-literal positive clause
stating that the term corresponding to the cell must equal
one of the domain members.  For example,
\[
f(2,3) = 0 \ \ | \ \ f(2,3) = 1 \ \ | \ \ f(2,3) = 2 \ \ | \ \ f(2,3) = 3.
\]
Now we have a satisfiability problem for domain size $n$,
independent of Mace4.

Practical decision procedures for 
ground clauses with equality have been studied before.
However, the structure of our problems leads us to believe
that the special-purpose methods embodied in Mace4 are better
than general-purpose methods that have been proposed.  First,
the $n$-literal positive clauses indicate that the
splitting and backtracking approach in Mace4 is appropriate.
Second, it is not clear how the symmetry exploited by the
least-number heuristic could be effectively addressed in a general-purpose
method.  Third, rewriting with assignment clauses is
the only kind of rewriting we need, because we can always get
enough assignment clauses (by splitting on the
$n$-literal positive clauses) to rewrite every clause to 0 or 1.
Finally, negative inference is not needed for completeness,
but it frequently gives enormous speedups.

Before designing Mace4, we discussed the problem with
N. Shankar, R. Niewenhuis, and D. Kapur.
Most of those discussions revolved around combining congruence
closure (to do the ground rewriting) with the
DPLL method (to do the splitting and backtracking).
That is still an interesting approach, but we took
the special-purpose path, because it
seems simpler and more natural for this application.

\section{Auxiliary Programs}  \label{sec-aux}

The Mace4 distribution packages contain several
additional programs and scripts.  Most of these are
short C programs built from the routines in LADR library,
and they take inputs similar to the Mace4 input.

\paragraph{Isofilter.}  The least number heuristic in Mace4
prevents some of the isomorphism in the search for models,
but it is far from optimal.  If the user asks for
more than one model for a particular input
(e.g., \texttt{assign(max\_models,10000)},
most of the models will usually be isomorphic to
others in the set.  The program isofilter
takes a stream of models and eliminates the
isomorphic ones.  Let's say we wish to see all of
the ortholattices (OLs) up through order 10.
We prepare the input
following file, named \texttt{OL.in}.
\begin{center}
\begin{small}
\begin{boxedverbatim}
op(400, infix, ^).  op(400, infix, v).
assign(iterate_up_to, 10).
set(print_models_portable).
assign(max_models, 100000).

clauses(theory).
%
    x v y = y v x.              x ^ y = y ^ x.
    (x v y) v z = x v (y v z).  (x ^ y) ^ z = x ^ (y ^ z).
    x v (x ^ y) = x.            x ^ (x v y) = x.
%
    x v c(x) = 1.
    x ^ c(x) = 0.
%
    c(x ^ y) = c(x) v c(y).
    c(x v y) = c(x) ^ c(y).
    c(c(x)) = x.
%
%
%
%
%
%
end_of_list.
\end{boxedverbatim}
\end{small}
\end{center}
Then we run the following command.

\begin{small}
\begin{verbatim}
    %
\end{verbatim}
\end{small}
The output file \texttt{OL.out} contains 24 ortholattices of orders 2--10,
and the last line of the file is

\begin{small}
\begin{verbatim}
isofilter: input=1315, kept=24, checks=1291, perms=2975947, 2.27 sec.
\end{verbatim}
\end{small}
saying that isofilter received 1,315 interpretations from Mace4
and determined that
24 of those are nonisomorphic.  It did 1,291 actual isomorphism checks
on pairs of interpretations, involving almost 3 million permutations.
Generating the OLs took about 1.4 seconds (this time was determined
in a separate run, because the script \texttt{get\_interps} removed
that statistic), and eliminating the isomorphic ones took 2.27 seconds.

Isofilter does not attempt to permute operations when checking
for isomorphism.  For example, a pair of dual lattices are
not necessarily isomorphic.

\paragraph{Get\_interps.}
This is a simple awk script that extracts interpretations from
Mace4 output.  The interpretations must be in \emph{portable format},
which is specified with the Mace4 command
\texttt{set(print\_models\_portable)}.  See the preceding
\texttt{isofilter} example.

\paragraph{Modfilter.}
This program uses a set of interpretations to filter a stream of clauses.
It takes two command-line arguments: (1) a file of interpretations
in portable format, and (2) the type of filter to apply.  The
filter types are listed in the following table.
\begin{center}
\begin{tabular}{ll}
\hline
\texttt{true\_in\_all} & Admit clauses true in all of the interpretations.\\
\texttt{true\_in\_some} & Admit clauses true in some of the interpretations.\\
\texttt{false\_in\_all} & Admit clauses false in all of the interpretations.\\
\texttt{false\_in\_some} & Admit clauses false in some of the interpretations.\\
\hline
\end{tabular}
\end{center}
The clauses to be tested are read from the standard input, which
should contain nothing but the clauses (without \texttt{end\_of\_list}).
If the clauses need operator declarations, they should
be at the beginning of the file of interpretations.
The interpretations should be surrounded by 
``\texttt{terms(interpretations).}'' and ``\texttt{end\_of\_list.}''.
Here is an example that uses
the ortholattices up through size 6 to filter some Boolean algebra
identities.

\begin{small}
\begin{verbatim}
    %
\end{verbatim}
\end{small}

\paragraph{Modtester.}
This program takes a set of interpretations and a stream of clauses.
For each clause, it tells the interpretations in which the clause is true.
The calling sequence and the format of the input are similar to
the program modfilter, except that no filter type is given on the
command line.  An example follows.

\begin{small}
\begin{verbatim}
    %
\end{verbatim}
\end{small}

\paragraph{Interpfilter.}
This program takes a file of clauses (as command-line argument 1),
a test to apply (as command-line argument 2),
and a stream of interpretations (from the standard input).
The accepted tests are \texttt{models} and \texttt{nonmodels}.
For each interpretation I, if I is a model or nonmodel
(as specified in the test) of the set of clauses,
the interpretation is passed through to the standard output.
Say we have a set of ortholattices (file \texttt{OL.8})
and we wish to eliminate the distributive ones.
Then we can create a file containing a distributivity clause
(file \texttt{distributivity}) and run the following command.

\begin{small}
\begin{verbatim}
    %
\end{verbatim}
\end{small}

\paragraph{Otter-to-mace4.}
This is a Perl script that takes an Otter (also Mace2) input
file (from the standard input) and tries to convert it to a Mace4
input file.  It is far from perfect, but is works well
for many simple Otter inputs.  An example follows.

\begin{small}
\begin{verbatim}
    %
\end{verbatim}
\end{small}

\noindent 
For more information, run the command ``\texttt{otter-to-mace4 help}''.

\addcontentsline{toc}{section}{References}

\raggedright  %

\bibliographystyle{plain}

\end{document}